\documentclass[conference]{IEEEtran}

\ifCLASSINFOpdf
\else
\fi
\usepackage{booktabs}
\usepackage{amsmath,amssymb,amsfonts}
\usepackage{algorithmic}
\usepackage{graphicx}
\usepackage{textcomp}
\usepackage{xcolor}
\usepackage{hyperref}
\usepackage{array}
\usepackage{siunitx}
\usepackage{booktabs}
\usepackage{float}
\usepackage{hyperref}
\usepackage{listings} % For code listing
\usepackage{color} % For colouring code
\usepackage{xcolor}
\usepackage{hyperref}
\definecolor{codegreen}{rgb}{0,0.6,0}
\definecolor{codegray}{rgb}{0.5,0.5,0.5}
\definecolor{codepurple}{rgb}{0.58,0,0.82}
\definecolor{backcolour}{rgb}{0.95,0.95,0.92}
\definecolor{codeblue}{rgb}{0.25, 0.5, 0.75}

\usepackage{listings}
\usepackage{xcolor}
\usepackage{array}

\lstdefinestyle{mystyle2}{
    backgroundcolor=\color{backcolour},   
    commentstyle=\color{green},
    keywordstyle=\color{blue},
    stringstyle=\color{purple},
    basicstyle=\ttfamily\footnotesize, % Reduce font size
    breakatwhitespace=false,         
    breaklines=true,                 
    captionpos=b,                    
    keepspaces=true,                 
    numbers=none,                                  
    showspaces=false,                
    showstringspaces=false,
    showtabs=false,                  
    tabsize=1
}

\lstdefinestyle{mystyle}{
    backgroundcolor=\color{backcolour},   
    commentstyle=\color{codeblue},
    keywordstyle=\color{magenta},
    stringstyle=\color{codepurple},
    basicstyle=\ttfamily\footnotesize,
    breakatwhitespace=false,         
    breaklines=true,                 
    captionpos=b,                    
    keepspaces=true,                 
    numbers= none,                           
    showspaces=false,                
    showstringspaces=false,
    showtabs=false,                  
    tabsize=1
}

\begin{document}
%
% paper title
% Titles are generally capitalized except for words such as a, an, and, as,
% at, but, by, for, in, nor, of, on, or, the, to and up, which are usually
% not capitalized unless they are the first or last word of the title.
% Linebreaks \\ can be used within to get better formatting as desired.
% Do not put math or special symbols in the title.
\title{TD-Suite: All Batteries Included Framework for Technical Debt Classification }

% author names and affiliations
% use a multiple column layout for up to three different
% affiliations
\author{
    \IEEEauthorblockN{Karthik Shivashankar}
    \IEEEauthorblockA{University of Oslo\\
    karths@ifi.uio.no}
    \and
        \IEEEauthorblockN{Antonio Martini}
    \IEEEauthorblockA{University of Oslo\\
    antonima@ifi.uio.no}
}

% use for special paper notices
%\IEEEspecialpapernotice{(Invited Paper)}

% make the title area
\maketitle

% As a general rule, do not put math, special symbols or citations
% in the abstract
\begin{abstract}

Recognizing that technical debt is a persistent and significant challenge requiring sophisticated management tools, TD-Suite offers a comprehensive software framework specifically engineered to automate the complex task of its classification within software projects. It leverages the advanced natural language understanding of state-of-the-art transformer models to analyze textual artifacts, such as developer discussions in issue reports, where subtle indicators of debt often lie hidden.

TD-Suite provides a seamless end-to-end pipeline, managing everything from initial data ingestion and rigorous preprocessing to model training, thorough evaluation, and final inference. This allows it to support both straightforward binary classification (debt or no debt) and more valuable, identifying specific categories like code, design, or documentation debt, thus enabling more targeted management strategies.

To ensure the generated models are robust and perform reliably on real-world, often imbalanced, datasets, TD-Suite incorporates critical training methodologies: k-fold cross-validation assesses generalization capability, early stopping mechanisms prevent overfitting to the training data, and class weighting strategies effectively address skewed data distributions. Beyond core functionality, and acknowledging the growing importance of sustainability, the framework integrates tracking and reporting of carbon emissions associated with the computationally intensive model training process.

It also features a user-friendly Gradio web interface in a Docker container setup, simplifying model interaction, evaluation, and inference.
\end{abstract}
% no keywords

\IEEEpeerreviewmaketitle

\section{Introduction}
The effective management of technical debt stands as a critical, yet often underestimated, challenge within the landscape of modern software development. Technical debt, metaphorically representing the accumulated cost of future rework stemming from expedient, short-term design or implementation choices over more optimal, sustainable solutions, exerts a profound influence on software quality, long-term maintainability, system evolution, and overall team productivity  \cite{buschmann2011pay}. As software systems inevitably grow in scale and complexity over their lifecycle, the manual identification and assessment of technical debt through conventional methods like code reviews and architectural audits become increasingly time-consuming, resource-intensive, and prone to inconsistency \cite{sutoyo2023self}. Recognizing these limitations, the software engineering community has increasingly turned towards automated approaches. Recent significant advancements in machine learning, particularly the emergence and refinement of sophisticated transformer-based language models such as BERT, RoBERTa \cite{devlin2019bert,liu2020roberta}, DistilRoBERTa and DeBERTaV3 \cite{sanh2019distilbert,he2023debertav}, present highly promising avenues for automating the nuanced tasks of detecting and classifying technical debt embedded within various software artifacts, including source code comments, commit messages, issue tracker descriptions, and requirements documentation.

Despite the demonstrated potential of these advanced machine learning techniques in research contexts, a discernible gap persists between theoretical advancements and their practical, widespread adoption in industrial software development environments \cite{Li2015Mapping, sutoyo2023self}. Many existing tools and proposed solutions often focus narrowly on specific types of technical debt, lack seamless integration capabilities with standard development workflows, or require substantial specialized expertise for effective configuration and utilization \cite{ernst2021technical, sutoyo2023self}. Consequently, there is a compelling need for a comprehensive, integrated framework that not only harnesses the analytical power of state-of-the-art transformer models but also prioritizes practical usability, accessibility, and ease of deployment for software practitioners.

This paper introduces TD-Suite, a robust and extensible framework made for the automated Text classification of technical debt issues from source trackers like GitHub and also from Jira . The primary contributions of TD-Suite are multifaceted. Firstly, it provides a modular and cohesive architecture supporting both binary classification (identifying the presence or absence of debt) and categorizing debt into specific types like code debts, design debts, or documentation debts by ensembling different types of TD, leveraging transformer architectures for text classification task. Secondly, TD-Suite incorporates a comprehensive and flexible data processing pipeline capable of ingesting data from various sources and formats, including common structures like CSV files. Thirdly, the framework features an advanced training and evaluation system equipped with essential functionalities like k-fold cross-validation for robust performance estimation, early stopping to mitigate overfitting during training, mechanisms for handling class imbalance often present in technical debt datasets, and detailed performance metric reporting (e.g., accuracy, precision, recall, F1-score, MCC) complemented by insightful visualizations such as confusion matrices. Fourthly, and crucially for practical adoption, TD-Suite offers user-friendly deployment options, including an interactive web interface built with Gradio that allows users to fine-tune models and evaluate performance without deep technical knowledge, and a containerized deployment solution using Docker, ensuring environmental consistency and simplifying integration into diverse development and operational settings. By consolidating these capabilities, TD-Suite empowers software engineering teams to more effectively identify, understand, and manage technical debt.

\section{Background}

\subsection{Natural Language Processing (NLP) and Transformers for Text Classification}
NLP increasingly relies on machine learning  rather than rule-based methods \cite{kowsari2019text} for tasks like text classification \cite{aggarwal2012survey}, where text is automatically assigned predefined labels. This powers applications ranging from spam filtering \cite{sharma2023comparative} and sentiment analysis \cite{medhat2014sentiment} to identifying TD within software engineering artifacts like commit messages or issue descriptions \cite{maldonado2017detecting} by training models on manually labeled examples. Despite challenges posed by language ambiguity \cite{allamanis2018survey} and large data volumes, the development of the Transformer architecture marked a significant advancement.

Transformers utilize a self-attention mechanism \cite{vaswani2017attention} to effectively weigh word importance and capture long-range contextual dependencies, outperforming previous models. This innovation facilitated the creation of large Pre-trained Language Models (PLMs) such as BERT \cite{devlin2019bert}, which learn general language understanding from vast text corpora. These powerful PLMs, often made accessible through platforms like Hugging Face \cite{wolf2019huggingface}, can then be efficiently fine-tuned on smaller, task-specific datasets (e.g., TD-annotated text) to achieve high performance in classification and other NLP applications \cite{howard2018universal, chernyavskiy2021transformers}.

\section{Related Work}
The systematic detection and classification of technical debt within software projects has been an active research area for over a decade, evolving significantly with advances in software analysis and machine learning \cite{sutoyo2023self}. Initial efforts predominantly relied on static code analysis tools and predefined rule-based systems \cite{tufano2017empirical}, focusing on identifying coding standard violations, detecting specific code smells (e.g., long methods, large classes), or calculating complexity metrics (e.g., cyclomatic complexity). While valuable for localized structural issues, these early methods often lacked contextual understanding, struggled to differentiate intentional design from genuine debt, generated numerous findings requiring manual filtering, and were largely confined to source code analysis, neglecting other crucial information sources like documentation. Consequently, they provided an incomplete picture, often failing to capture systemic or non-code-related debt forms like architectural mismatches or inadequate testing \cite{gu2024self}.

As machine learning matured, researchers explored its application to technical debt identification, moving beyond simple rule matching \cite{Huang2018TextMining}. Traditional models like Support Vector Machines (SVMs), Naive Bayes, and Random Forests were trained on labeled datasets (code snippets, commit messages) to predict debt presence or type. These methods improved over static analysis by learning patterns, yet often relied on handcrafted features from text or code metrics, potentially missing complex semantic nuances indicative of certain debt types. Their effectiveness heavily depended on the quality and representativeness of the feature engineering \cite{Tsoukalas2021Compare,Li2015Mapping}.

The advent of deep learning, particularly transformer models like BERT and its variants \cite{devlin2019bert}, shifted the paradigm in natural language processing and subsequently influenced software engineering research \cite{feng2020codebert}.

Research into automated Technical Debt (TD) and Self-Admitted Technical Debt (SATD) identification has progressed from early Natural Language Processing (NLP) techniques for classifying SATD types \cite{Maldonado2017NLP} and N-gram based approaches \cite{Wattanakriengkrai2018Ngram}, to exploring word embeddings like Word2Vec \cite{Flisar2018Word2Vec}. Deep Learning models were subsequently applied, including Convolutional Neural Networks (CNNs) for SATD detection in Infrastructure as Code \cite{Ren2019CNN} and alongside Recurrent Neural Networks (RNNs) for analyzing SATD removal \cite{Zampetti2020CNN_RNN}. LSTMs were also utilized for SATD type classification \cite{Santos2020LSTM}, and various studies compared traditional ML with DL models like CNNs, RNNs, and LSTMs \cite{Tsoukalas2021Compare, Li2023TextCNN, Alhefdhi2022DL}. 

More recently, Transformer architectures have demonstrated strong results; BERT was found effective for TD detection in R package reviews \cite{Khan2022BERT}, while Transformers have been applied to classify TD \cite{Skryseth2023Transformers} and identify TD types \cite{shivashankar2025beacon} in issue trackers, with recent reviews corroborating BERT's effectiveness \cite{Melin2024Review}. This body of work relies on curated datasets derived from sources such as Apache projects \cite{Lenarduzzi2019Dataset}, R package reviews and code comments \cite{Codabux2021R_Review, Vidoni2021R_SatD}, and ML software \cite{OBrien2022ML_SATD}.

Surveying the current landscape reveals a fragmented tool ecosystem \cite{ajibode2024systematic}. Existing solutions often specialize: some in static analysis for code-level debt, others in architectural metrics, and others using NLP on issue trackers or documentation. Few tools offer a holistic approach identifying and classifying multiple debt types (code, design, documentation, test, etc.) in one integrated platform. This fragmentation challenges teams seeking a comprehensive view of project health \cite{shivashankar2025beacon}.

Furthermore, critical limitations persist. A primary issue is lacking comprehensive coverage, with most tools targeting only a subset of the technical debt spectrum. Integrating multiple specialized tools is often complex, burdensome, and hinders adoption \cite{ernst2021technical}. Evaluation practices vary considerably; non-standardized datasets and metrics make reliable comparison of tools difficult. Finally, the computational resources and environmental impact (e.g., carbon emissions) of sophisticated models are often overlooked aspects gaining importance for sustainable software development. TD-Suite is deliberately designed to address these shortcomings by providing a unified, extensible framework covering multiple debt categories, emphasizing easy integration (Docker) and accessible interfaces (Gradio CLI/Web), employing rigorous evaluation, and incorporating resource consumption awareness, offering a more complete and practical solution.

\section{TD-Suite Framework}
\begin{figure}
    \centering
    \includegraphics[width=1\linewidth]{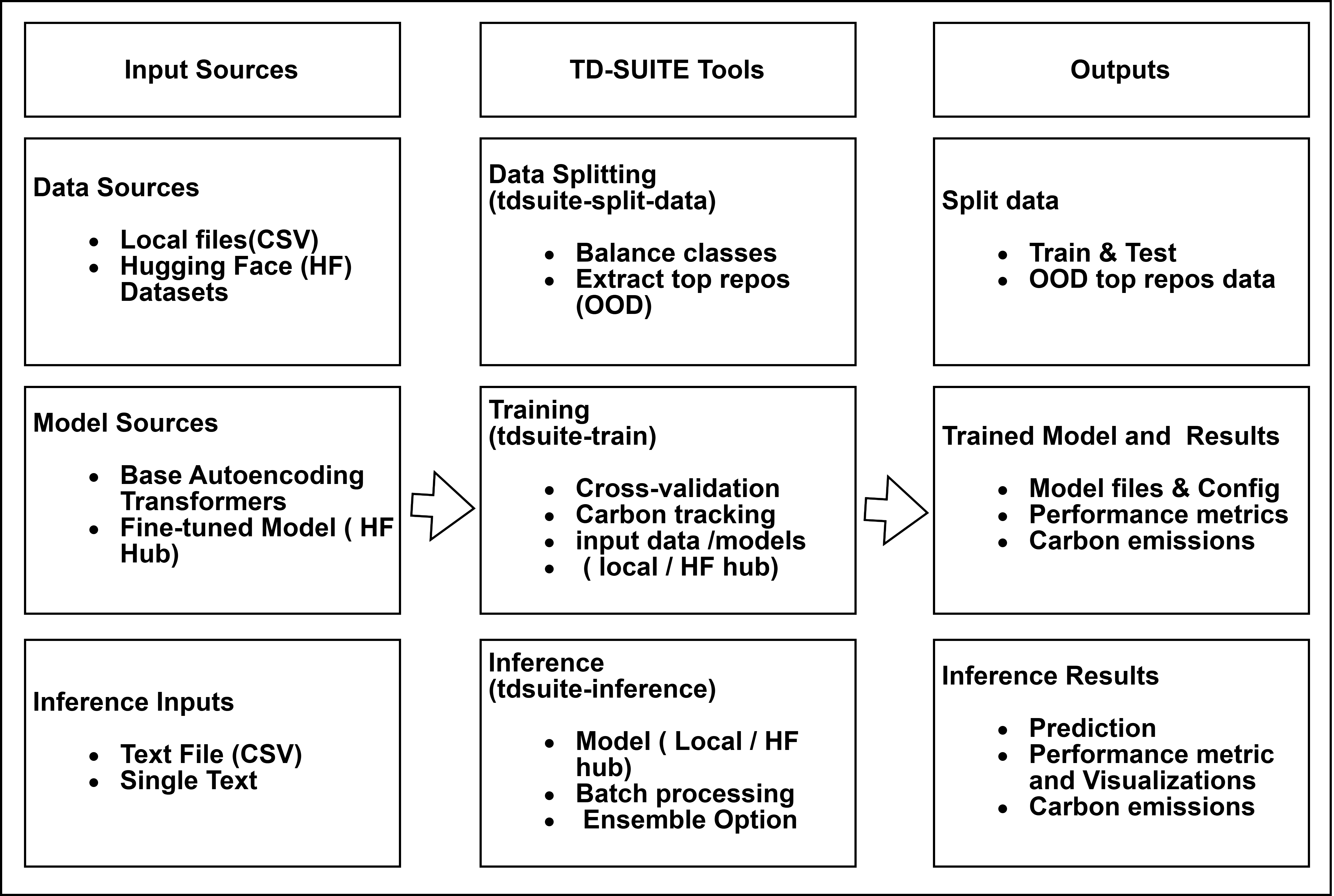}
    \caption{TD-Suite High level Architecture }
    \label{fig:TDSuite}
\end{figure}

The TD-Suite framework is architecturally designed with modularity and extensibility as core principles, aiming to strike a balance between high classification performance, practical usability for software engineers, and efficient resource utilization. Its structure is organized into interconnected components, each responsible for a distinct stage of the technical debt text classification workflow. This modular design facilitates maintainability, allows for independent upgrades of components, and supports future extensions to accommodate new models or data processing techniques. These core components collaboratively deliver a comprehensive solution, guiding the process from initial data input through to model application and result interpretation as shown in Figure \ref{fig:TDSuite}.

The first major component is the Data Management layer, which serves as the foundation for the entire classification process. This layer is responsible for the ingestion, validation, processing, and preparation of data intended for model training or inference. It is engineered to handle diverse data sources and formats commonly encountered in software development, including structured files like Comma Separated Values (CSV) and online repositories such as Hugging Face Datasets. Upon ingestion, the data undergoes rigorous validation to ensure integrity and adherence to expected schemas, typically requiring 'text' and 'label' fields. Subsequently, a customizable preprocessing pipeline is applied, encompassing essential steps like text cleaning , normalization (e.g., lowercasing), potentially stemming or lemmatization, and crucially, the transformation of raw text into numerical representations suitable for transformer models via tokenization. This component also manages the critical task of data splitting, partitioning datasets into training, validation, and test sets using strategies like stratified sampling to maintain class distribution. The processed and tokenized data is then structured into efficient dataset objects, often leveraging libraries like Hugging Face's `datasets` \cite{huggingface_datasets}, ready for consumption by the subsequent model and training layers.

The Training Pipeline component orchestrates the complex process of training the selected transformer models on the prepared datasets. It leverages high-level training libraries like the Hugging Face Trainer API \cite{huggingface_transformers},  to manage the training loop, optimization, and evaluation cycles. Beyond basic training, this pipeline integrates several advanced features crucial for developing robust and reliable models in the often-challenging context of technical debt data. It supports k-fold cross-validation, enabling a more rigorous assessment of model generalization by training and evaluating on multiple subsets of the data. Early stopping mechanisms are incorporated to monitor validation performance during training and halt the process when improvement ceases, effectively preventing overfitting and saving computational resources. Recognizing that technical debt datasets are frequently imbalanced (with non-debt instances often outnumbering debt instances, or certain debt types being rare), the pipeline provides options for applying class weighting techniques during loss calculation, ensuring that the model pays adequate attention to minority classes. Furthermore, this component includes capabilities for monitoring computational resource usage (CPU, GPU memory) and, significantly, tracking the associated carbon emissions generated during the computationally intensive training phase, promoting environmentally conscious model development \cite{anthony2020carbontracker}. 

Finally, once a model is trained, the Inference Engine component facilitates its practical application for classifying new, unseen data. This component is optimized for efficiency and scalability, supporting batch processing to classify large numbers of text instances simultaneously, thereby reducing computational overhead. It handles the loading of pre-trained model checkpoints and their associated tokenizers, performs the necessary preprocessing and tokenization on the input data, executes the model prediction, and formats the output, typically providing both the predicted class label and associated confidence scores or probabilities. The inference process, can operate on individual text inputs or entire datasets provided in compatible formats. The design also allows for potential extensions towards ensemble methods, where predictions from multiple models could be combined to potentially improve overall accuracy and robustness \cite{shivashankar2025beacon}. Together, these components form a cohesive and powerful framework for tackling the multifaceted challenge of technical debt classification in software engineering. We have open sourced this package with comprehensive documentation and evaluation  \footnote{\url{https://github.com/KarthikShivasankar/text\_classification}}.

\section{Results}
To assess the capabilities and practical effectiveness of the TD-Suite framework, a comprehensive set of experiments was designed and executed. This evaluation aimed to quantify the framework's performance on core technical debt classification tasks  from SATD Github isssues datasets \cite{shivashankar2025beacon}, measure its efficiency in terms of computational resource consumption, and validate the robustness of different BERT based models . The experimental design encompassed multiple  representative transformer model architectures, and standardized evaluation metrics. 

\subsection{Experimental Setup}
The foundation of our evaluation rested upon carefully selected datasets representing diverse scenarios encountered in technical debt analysis. We utilized TD  Datasets and its different Types like Code , Documentation , Testing,  Architecture debts and so on  from  K Shivashankar et.al \cite{shivashankar2025beacon}. This dataset was balanced, ensuring an equal representation of both positive and negative classes, providing a solid baseline for assessing fundamental debt identification capabilities. 

For the model evaluation, we selected four widely recognized and performant transformer architectures, representing different points on the performance-efficiency spectrum. These included BERT-base-uncased, a foundational and extensively benchmarked model; RoBERTa-base, known for its optimized pre-training procedure leading to often superior performance \cite{devlin2019bert,liu2020roberta}; and DistilRoBERTa-base-uncased, a distilled version offering significantly faster inference and reduced memory footprint while retaining substantial performance and finally DeBERtav3-base \cite{sanh2019distilbert,he2023debertav} . 

To ensure fair comparison and reproducibility, a consistent set of hyperparameters was applied across all experiments involving these models unless otherwise specified. The maximum input sequence length was set to 512 tokens, accommodating the typically verbose nature of software artifacts. A batch size of 32 was used during training, balancing computational efficiency and gradient stability. The learning rate was set to the commonly used value of 2e-5 for fine-tuning transformers, and models were trained for a standard duration of 3 epochs. A warmup phase consisting of 500 steps was employed to stabilize training in the initial stages. These standardized configurations allowed for a direct comparison of the models' inherent capabilities within the TD-Suite framework.

 We assessed model performance using standard metrics such as accuracy, precision, recall, Matthews’s correlation coefficient (MCC), and F1-score on a reserved test set. 
Precision measures the percentage of TD issues that have been correctly identified (true positives) out of all predictions.
The Recall metric, which calculates the proportion of correctly identified TD issues (true positives) out of all actual TD issues (true positives + false negatives), holds particular importance in our study.

We also emphasize Matthews’s Correlation Coefficient (MCC), which is widely recognized as a more statistically reliable metric \cite{chicco_advantages_2020} for binary classification. 

{\scriptsize
\[
MCC = \frac{TP \cdot TN - FP \cdot FN}{\sqrt{(TP + FP) \cdot (TP + FN) \cdot (TN + FP) \cdot (TN + FN)}}
\]
}
MCC produces a more informative and truthful score in evaluating classifications than accuracy and F1 score. It produces a high score only if the prediction obtained good results in all four confusion matrix categories (true positives (TP), false negatives (FN), true negatives (TN), and false positives (FP)), proportionally both to the size of positive and negative elements in the dataset.
 MCC values range between 1 to -1, and a value greater than 0 indicates that the classifier is better than a random flip of a fair coin, whilst 0 means it is no better. 

In addition, all models were also tested against an test split of the dataset, which consisted of data from projects not included in the training dataset. This step is crucial to determining the models' generalization capabilities to new, unseen data contexts.

\subsection{Data availability}
We have made the replication package publicly available at \href{https://doi.org/10.5281/zenodo.15161451}{\textbf{ https://doi.org/10.5281/zenodo.15161451}} \footnote{\url{https://doi.org/10.5281/zenodo.15161451}}.

Additional resources related to the TD-suite package, models, and datasets include:
\begin{itemize}
    \item \textbf{TD-suite GitHub Project:} The source code and documentation are available at \href{https://github.com/KarthikShivasankar/text_classification}{https://github.com/KarthikShivasankar/text\_classification}.
    \item \textbf{TD-suite Models:} A collection of models for use with TD-suite can be found on Hugging Face at \href{https://huggingface.co/collections/karths/technical-debts-and-its-types-text-classification-models-67f91939c82b2f6081e06cd8}{TD-suite Model Collections Hyperlink}.
    \item \textbf{TD-suite Datasets:} Associated datasets are available in a Hugging Face collection at \href{https://huggingface.co/collections/karths/technical-debt-and-it-types-datasets-67f91c2657176ce47372fab6}{TD-suite Dataset Collections Hyperlink}.
\end{itemize}

\subsection{Performance Metrics obtained from TD-Suite framework}

\begin{table}[htbp]
  \centering
 
  \caption{Comparison of Performance Metrics}
  \label{tab:performance_comparison} % New label
  \begin{tabular}{l r r r r}
    \toprule
    \textbf{Metric} & \textbf{BERT} & \textbf{RoBERTa} & \textbf{DistilRoBERTa} & \textbf{DeBERTaV3} \\
    \midrule
    Accuracy          & 0.8831 & 0.8827 & 0.8857 & 0.8983 \\ % Added DeBERTaV3 data
    Precision         & 0.9174 & 0.8254 & 0.8389 & 0.8635 \\ % Added DeBERTaV3 data
    Recall            & 0.8013 & 0.9236 & 0.9099 & 0.9078 \\ % Added DeBERTaV3 data
    F1 Score          & 0.8554 & 0.8717 & 0.8730 & 0.8851 \\ % Added DeBERTaV3 data
    MCC               & 0.7630 & 0.7684 & 0.7716 & 0.7947 \\
    \bottomrule
  \end{tabular}
\end{table}

Table \ref{tab:performance_comparison} presents a comparison of the performance metrics for the BERT, RoBERTa \cite{devlin2019bert,liu2020roberta}, DistilRoBERTa, and DeBERTaV3 models in classifying TD  \cite{sanh2019distilbert,he2023debertav}. 

We trained all these model on the train split of the TD datasets \cite{shivashankar2025beacon}. DeBERTaV3 achieves the best overall performance, with the highest accuracy (0.8983), precision (0.8635), and F1 score (0.8851). While RoBERTa has the highest recall (0.9236), it lags behind DeBERTaV3 in other metrics. DistilRoBERTa, a distilled version of RoBERTa, shows competitive performance, demonstrating the effectiveness of knowledge distillation in maintaining accuracy while reducing model size. BERT, while still strong, generally exhibits the lowest performance across these metrics. The Matthew's Correlation Coefficient (MCC) further supports these observations, with DeBERTaV3 achieving the highest MCC of 0.7947, indicating superior performance in binary classification .
These models trained using TD-Suite perform comparatively better in all classification performance metrics compared to other recent Deep Learning  models like CNNs, RNNs, and LSTMs \cite{Tsoukalas2021Compare, Li2023TextCNN, Alhefdhi2022DL}  for TD  text classification task.

\subsection{Resource Usage and Carbon Emissions}
\begin{table}[htbp]
  \centering
  \caption{Comparison of Resource Usage and Carbon Emissions}
  \label{tab:resource_comparison}
  \footnotesize % Use smaller font size
  \setlength{\tabcolsep}{4pt} % Reduce space between columns (default is 6pt)
  \begin{tabular}{l r r r r}
    \toprule
    % Slightly shortened headers if needed, but try without first
    \textbf{Parameter} & \textbf{BERT} & \textbf{RoBERTa} & \textbf{DistilRoB.} & \textbf{DeBERTaV3} \\
    \midrule
    \multicolumn{5}{l}{\textit{Inference}} \\
    Emissions (kgCO$_2$e) & 0.00103 & 0.00109 & 0.00059 & 0.00158 \\ % Used $_2$ instead of \textsubscript
    Duration (s)      & 143.46 & 151.72 &  81.99 & 215.23  \\
    \midrule % Separates Inference and Training sections
    \multicolumn{5}{l}{\textit{Training}} \\
    Emissions (kgCO$_2$e)  & 0.03677 & 0.03710 & 0.01904 & 0.05566  \\
    Duration (s)      & 5140.70 & 5183.37 & 2663.26 &  7674.37 \\
    \bottomrule
  \end{tabular}
\end{table}

Table \ref{tab:resource_comparison} provides a comparison of the resource usage and carbon emissions for the four models during both inference and training obtained from TD-Suite framework when Training in the train split  and running inference on the test split of the TD Dataset \cite{shivashankar2025beacon}. DistilRoBERTa demonstrates the lowest resource consumption, with the shortest inference duration (81.99 s) and the lowest training emissions (0.01904 kgCO\textsubscript{2}e). In contrast, DeBERTaV3, while achieving the best performance, exhibits the highest inference duration (215.23 s) and training emissions (0.05566 kgCO\textsubscript{2}e). BERT and RoBERTa show similar resource usage, with RoBERTa having slightly higher emissions and duration than BERT. These results highlight the trade-off between model performance and computational cost. Distillation, as seen with DistilRoBERTa, offers an effective strategy for reducing resource usage and environmental impact with almost comparable performance as shown in Table \ref{tab:performance_comparison}.

\subsection{Identify different Technical Debt Types}
\begin{table}[h]
\centering
\caption{Performance of Binary Classification (DistilRoBERTa)  Models on Different Technical Debt Types}
\label{tab:td_performance_condensed} % Changed label slightly for clarity
\begin{tabular}{ccccr} % Removed the first 'l' column specifier
\hline
 Recall & Accuracy & F1 & Model & Support \\ % Removed 'TD Type &' header
\hline
 0.825 & 0.825 & 0.904 & Test & 1950 \\ % Removed 'Test &'
 0.972 & 0.972 & 0.986 & Technical Debt & 1950 \\ % Removed 'Test &'
\hline
 0.755 & 0.755 & 0.860 & Architecture & 49 \\ % Removed 'Architecture &'
 0.980 & 0.980 & 0.990 & Technical Debt & 49 \\ % Removed 'Architecture &'
\hline
 0.736 & 0.736 & 0.848 & Code & 469 \\ % Removed 'Code &'
 0.970 & 0.970 & 0.985 & Technical Debt & 469 \\ % Removed 'Code &'
\hline
 0.856 & 0.856 & 0.922 & Documentation & 528 \\ % Removed 'Documentation &'
 0.930 & 0.930 & 0.964 & Technical Debt & 528 \\ % Removed 'Documentation &'
\hline
\end{tabular}
\end{table}
Table~\ref{tab:td_performance_condensed} presents the performance results for binary classification models  , specifically DistilRoBERTa, tasked with identifying different types of technical debt (TD) using TD-Suite framework on the test split unseen by the model during training. The table contrasts the performance of models trained for specific TD types against a general TD detection model across four categories: Test, Architecture, Code, and Documentation Debt using Github SATD  issues  Dataset from  K Shivashankar et.al containing all these different TD issues types and more \cite{shivashankar2025beacon} .

For each category, two rows of metrics (Recall, Accuracy, F1 Score) are provided, along with the support count (number of samples). The first row within each category (labeled 'Test', 'Architecture', 'Code', or 'Documentation') reflects the performance of a binary classifier trained specifically to identify that particular type of technical debt against other types or non-debt instances. The performance for these type-specific classifiers shows moderate success, with Recall and Accuracy values ranging from 0.736 (for Code Debt) up to 0.856 (for Documentation Debt), and corresponding F1 scores between 0.848 and 0.922.

The second row within each category (consistently labeled 'Technical Debt') shows the performance of a general binary classifier trained simply to detect the presence of \textit{any} technical debt, evaluated on the same set of samples pertaining to that specific category. These results demonstrate significantly higher performance across all metrics. Recall and Accuracy are consistently high, ranging from 0.930 to 0.980, and F1 scores are correspondingly strong, ranging from 0.964 to 0.990.

The primary conclusion drawn from this table is that while identifying specific types of technical debt using binary classification is feasible to a reasonable degree, a general binary classifier achieves substantially higher accuracy and recall in simply determining whether technical debt is present or not. This suggests that the nuanced differences between specific TD types are harder for the model to distinguish compared to the broader distinction between the presence and absence of TD. 

Given the observed poor performance of multi-class classifiers (as indicated in related work like BEACon-TD) \cite{shivashankar2025beacon} and the varying performance of the binary classifiers shown in Table~\ref{tab:td_performance_condensed}, a "TD Suite" approach using an ensemble of two distinct binary classification stages is proposed to effectively identify different TD types. Here's how it would work:

\paragraph{Initial TD Detection:} The first stage uses the highly effective general binary classifier (the "Technical Debt" model in Table~\ref{tab:td_performance_condensed}, showing high recall/accuracy). This model acts as a gatekeeper, accurately identifying whether a given software issue or text snippet contains \textit{any} form of technical debt. Its high recall ensures that most instances of TD are captured.

\paragraph{Specific TD Type Classification:} If the first model positively identifies the presence of TD, a second stage is triggered. This stage employs the "Binary types" classifiers (either one model trained to distinguish types or a set of models, each specialized for one type, similar to those evaluated in the first row for each category in Table~\ref{tab:td_performance_condensed}). These models, while having lower performance than the general TD detector, are better at pinpointing the specific category (e.g., Code Debt, Test Debt, Documentation Debt) than a direct multi-class model.

\subsection*{Why this Ensemble Helps}

This two-stage ensemble approach offers several advantages:

\begin{itemize}
    \item \textbf{Leverages Strengths:} This approach leverages the strength of the general TD model (high detection accuracy) and the moderate strength of the specific TD type models (type identification).
    \item \textbf{Circumvents Multi-Class Failure:} It avoids the complexities and poor performance observed when trying to make a single model distinguish between many specific TD types simultaneously (the failed multi-class approach mentioned in the BEACon-TD context) \cite{shivashankar2025beacon}. By breaking the problem into two simpler binary classification steps (Is it TD? If yes, which type?), the overall classification accuracy and reliability for identifying specific TD types are improved compared to the direct multi-class method.
    \item \textbf{Efficiency:} The more complex type-classification models are only run on items already flagged as potential TD, potentially saving computational resources compared to running all type-specific classifiers on every single input.
\end{itemize}

TD Suite uses this two-stage ensemble approach to overcome the limitations of multi-class issues classification and provide a more robust mechanism for both detecting technical debt and classifying its specific type within diverse software project issues.

\section{Deployment and Usability Features}
Beyond the core classification capabilities and robust implementation, the practical success of a framework like TD-Suite hinges critically on its ease of deployment, accessibility, and overall usability for its intended audience: software engineers and researchers \cite{Li2015Mapping, ernst2021technical}. Recognizing this, TD-Suite incorporates several features specifically designed to streamline its integration into development workflows and make its powerful analytical tools readily available to users with potentially varying levels of machine learning expertise. These features span interactive interfaces and easy evaluation and testing of these different TD and its types models on custom datasets.

\subsection{Web Interface}
A key element enhancing TD-Suite's usability is its provision of an interactive web interface, implemented using the Gradio library. This interface serves as a graphical front-end, abstracting away much of the command-line complexity associated with model training and evaluation. It is structured logically into distinct tabs, primarily one for "Fine-tune Model" and another for "Evaluate Model," guiding the user through specific workflows. Within the fine-tuning tab, users can upload their technical debt datasets in common formats like CSV. The interface provides intuitive controls, such as sliders, for setting crucial preprocessing parameters, including the desired percentage of data to allocate for the training set (with the remainder typically used for testing the fine-tuned model) and a minimum word length threshold to filter out potentially uninformative short text instances. Upon initiating the "Process CSV" action, the application performs data loading, validation, cleaning, filtering, splitting, and saves the processed datasets into a structured local directory , providing immediate feedback on the status and displaying the head of the processed training data. Subsequently, users can select a base pre-trained model (e.g., a general TD classifier or a priority classifier) from a predefined dropdown list and initiate the fine-tuning process with a single button click. The interface displays the fine-tuning status and, upon completion, presents key performance metrics such as accuracy, a detailed classification report, a graphical confusion matrix plot, and a preview of the test dataset annotated with predictions and probabilities, making the results readily interpretable. Furthermore, it provides a link to download the full annotated dataset.

Similarly, the "Evaluate Model" tab allows users to upload a dataset specifically for evaluation purposes. Users can select one or multiple pre-trained or previously fine-tuned models  via checkboxes. After processing the input CSV , the user can trigger the evaluation process. The backend loads the selected models and the processed dataset, performs inference, potentially leveraging GPU acceleration if available, and aggregates the predictions. The interface then displays the head of the dataset augmented with prediction columns for each selected model and provides a downloadable spreadsheet containing the complete evaluation results. This Gradio application significantly lowers the barrier to entry for utilizing TD-Suite, enabling rapid experimentation, model fine-tuning on custom data, and straightforward evaluation without requiring users to write code or manage complex command-line arguments.

\subsection{Containerized Deployment}
To address the challenges often associated with setting up complex software environments, particularly those involving specific versions of machine learning libraries, system dependencies, and potentially CUDA drivers for GPU acceleration, TD-Suite provides a containerized deployment solution using Docker. The configuration is defined in the `Dockerfile` in our replication package from zenodo. This file specifies a suitable base image  that includes CUDA toolkit and cuDNN libraries, ensuring compatibility with NVIDIA GPUs for accelerated computation. The Dockerfile systematically installs necessary system packages.  It copies the TD-Suite application code into the container image and sets the default command to launch the Gradio web application.

This Docker-based approach offers several significant advantages. It guarantees a consistent and reproducible runtime environment across different machines (developer laptops, servers, cloud instances), eliminating "works on my machine" issues. Dependency management is encapsulated within the image build process, simplifying setup for end-users who only need Docker installed, thereby making the deployment process significantly more straightforward compared to manual environment configuration. This containerization strategy is crucial for promoting wider adoption and ensuring reliable operation in diverse IT infrastructures.

\subsection{Discussion}
In summary, the experimental evaluation robustly demonstrates that TD-Suite effectively addresses the intricate challenge of automated technical debt classification. The framework consistently achieves high levels of accuracy and reliability across  TD and its types datasets \cite{shivashankar2025beacon}, surpassing the performance typically associated with traditional ML methods and other recent DL models like CNNs, RNNs, and LSTMs \cite{Tsoukalas2021Compare, Li2023TextCNN, Alhefdhi2022DL}. The successful integration and evaluation of multiple transformer models highlight the framework's flexibility and provide users with clear options to balance performance requirements against resource constraints and sustainability goals. The cross-validation results further affirm the stability and generalizability of the models trained using the framework.

The combination of sophisticated transformer models, a meticulous data processing pipeline, advanced training features like cross-validation and early stopping contributes significantly to its overall effectiveness. This suggest that TD-Suite is not merely a theoretical construct but a practical, high-performing solution ready for application in real-world software engineering scenarios. Its demonstrated ability to handle diverse debt types and adapt to different contexts makes it a powerful tool for development teams seeking deeper insights into their projects' technical health and aiming to make more informed decisions regarding debt remediation and prevention. The quantifiable performance metrics and resource usage data provide a solid foundation for its adoption and integration into software development lifecycles.

\subsection{Impact and Significance}
TD-Suite offers substantial contributions to technical debt analysis through its robust and user-friendly framework. Key features include versatile data handling, advanced training techniques like cross-validation and early stopping to ensure model generalization and address class imbalance, and responsible AI practices via resource monitoring. Its dual interface (web UI and CLI) and Docker containerization ensure usability and easy deployment.

The framework significantly impacts both industry and research. Practitioners gain a tool for automated, data-driven technical debt identification and classification \cite{ernst2021technical, gu2024self}, improving strategic decisions on refactoring, resource allocation, and risk management for better code quality. Researchers benefit from a standardized platform for comparing models, experimenting with novel approaches, generating benchmarks, and fostering collaboration in the field of AI for software engineering challenges\cite{ernst2021technical, Li2015Mapping}.

\section{Threats to Validity}

Potential threats to construct validity primarily involve the measurement of TD. While standard classification metrics are used, accurately capturing the intended construct of specific TD subtypes (e.g., Code, Test, Architecture) proved challenging, as evidenced by lower model performance compared to general TD detection. This suggests potential ambiguity in the textual differentiation between subtypes, or limitations in the models' ability to discern these nuances. Furthermore, the validity relies on the consistency and accuracy of the initial expert labeling of the datasets, and the focus on textual artifacts means the measured construct may not encompass TD indicators present solely in non-textual sources.

Internal validity, concerning the causal relationship between the framework's components and the outcomes, appears reasonably controlled through the use of standardized hyperparameters for model comparison, cross-validation, early stopping, and methods to handle class imbalance. These techniques mitigate risks like overfitting and ensure fairer comparisons between different models' performance and resource consumption results. 

External validity, regarding the generalizability of findings, faces limitations. While the dataset drew from diverse repositories, the lack of detail on project characteristics (domain, language, scale) makes it uncertain how findings apply to different contexts. The study's focus on specific textual artifacts (like issue reports ) and the acknowledged need for future work exploring different languages, domains, and project types further highlight current constraints on generalizing the results.

\section{Conclusion}
In conclusion, this paper has introduced TD-Suite, a comprehensive and integrated software framework designed to address the persistent challenge of technical debt identification and classification using issues from source trackers within modern software development lifecycles . By  leveraging the sophisticated  transformer-based language models, TD-Suite provides a robust, automated approach to analyzing textual software artifacts and discerning various forms of technical debt. The framework successfully encapsulates an end-to-end workflow, encompassing data ingestion and preprocessing, flexible model selection and training, rigorous evaluation, and critically, user-friendly deployment mechanisms, thereby bridging a significant gap in practical software engineering application.

\bibliographystyle{IEEEtran}
\bibliography{sample-base}

% that's all folks
\end{document}